\documentclass[aps,prl,twocolumn,superscriptaddress,showpacs,showkeys]{revtex4-1}
\usepackage{graphicx}
\usepackage{graphics}
\usepackage{amsmath,amssymb,amsfonts}
\usepackage{colortbl}
\usepackage{times}
\usepackage{array}
\usepackage{bm}
\usepackage{soul}

\renewcommand*{\k}{\mathbf{k}}

\newcommand*{\Hc}{\textrm{H.c.}}

\newcommand*{\ra}{\rangle}
\newcommand*{\la}{\langle}
\newcommand*{\up}{\uparrow}
\newcommand*{\dn}{\downarrow}

\newcommand*\bsigma{\bm{\sigma}}
\newcommand{\cepd}{Ce$_3$Bi$_4$Pd$_3$}

\newcommand{\G}{\mathit{\Gamma}}

\begin{document}
\title{Extreme topological tunability of Weyl-Kondo semimetal to Zeeman coupling}
\author{Sarah E. Grefe}
\email{seg@lanl.gov}
\affiliation{Department of Physics and Astronomy and Rice Center for Quantum Materials, Rice University, Houston, Texas 77005, USA}
\affiliation{Theoretical Division, Los Alamos National Laboratory, Los Alamos, New Mexico 87545, USA}
\author{Hsin-Hua Lai}
\affiliation{Department of Physics and Astronomy and Rice Center for Quantum Materials, Rice University, Houston, Texas 77005, USA}
\author{Silke Paschen}
\affiliation{Institute of Solid State Physics, Vienna University of Technology, 1040 Vienna, Austria}

\author{Qimiao Si}
\affiliation{Department of Physics and Astronomy and Rice Center for Quantum Materials, Rice University, Houston, Texas 77005, USA}
\date{\small{\today}}
%%%%%%%%%%%%%%%%%%%%%%%%%%%%%%%%%%%%%%%%%%%%%%%%%%%%%%%%%%%%%%%%%%%%%%%%%%%%%%%%%%%%%%%%%%%%%%%%%
\begin{abstract}
	There is considerable interest in the intersection of correlations and topology, especially in metallic systems.
Among the outstanding questions are how strong correlations drive novel topological states and whether such states can be readily controlled.
	Here we study the effect of a Zeeman coupling on a Weyl-Kondo semimetal in a nonsymmorphic and noncentrosymmetric Kondo-lattice model. 
A sequence of distinct and topologically nontrivial semimetal regimes are found, each containing Kondo-driven and Fermi-energy-bound Weyl nodes. The nodes annihilate at a magnetic field that is smaller than what it takes to suppress the Kondo effect. As such, we demonstrate an extreme topological tunability that is isolated from the tuning of the strong correlations {\it per se}. Our results are important for experiments in strongly correlated systems, and set the stage for mapping out a global phase diagram for strongly correlated topology.
\end{abstract}
\maketitle
%
%%%%%%%%%%%%%%%%%%%%%%%%%%%%%%%%%%%%%%%%%%%%%%%%%%%%%%%%%%%%%%%%%%%%%%%%%%%%%%%%%%%%%%%%%%%%%%%%%%%%%%%%%%%%%%%%%%%%%%%%%

	The application of topological concepts to condensed matter physics has considerably enriched the landscape of quantum phases in the quantum matter field.
	Yet two crucial frontiers of the field remain: the search for tunable topological materials, and the exploration of topological states in strongly correlated settings.
	A plethora of quantum states of matter are driven by strong correlations~\cite{Keimer2017,Paschen2020}.
	Here, large Coulomb repulsion often produces local moments that form a building block of the low-energy physics.
	The quantum entanglement between the local moments determines the nature of the ground state.
	Of particular interest are heavy fermion systems, which possess both strong correlations and significant spin-orbit coupling that provides the band inversion suitable for nontrivial topology.

	In heavy fermion metals, itinerant electrons are quantum entangled with the local moments as well, giving rise to the celebrated Kondo effect~\cite{hewson_book} 
and a rich landscape of metallic quantum phases and quantum critical points~\cite{Si2020,Lohneysen2007}.
	When spin-orbit coupling is also strong, topological metallic states may also appear in the phase diagram.
	Recent theoretical~\cite{WKSM_PNAS,WKSM_PRB2020} and experimental~\cite{DzsaberPRL2017,Dzsaber2021} studies have shown that the Kondo effect can drive the emergence of Weyl nodes near the Fermi energy, leading to a Weyl-Kondo semimetal (WKSM).
	This development provides a new opportunity to utilize insights about the strong correlation physics of the heavy fermion metals and elucidate how strong correlations and topology intersect in metallic systems.

	Strong correlations {\it per se} amplify a system's response to external stimuli.
	For example, heavy fermion metals have long been explored by a magnetic field, which is a non-thermal tuning parameter and its relatively small variation can trigger phase transitions or help access quantum critical points by perturbing the $4f$ local moments~\cite{Heuser1998}.
	This is highlighted by the usage of a magnetic field to reveal a jump in the normal Hall effect that originates from an abrupt change in the Fermi surface at a local quantum critical point~\cite{paschen2004,Nair2012,SiPaschen2013,Shi05.1,Par06.1,Kne08.1,Si-Nature,Colemanetal,SI200623}.
	What has been open is whether strongly correlated topology can be readily tuned and, if so, what sequence
of topological quantum phase transitions can be realized.

	In this \textit{Letter}, we study the effect of a Zeeman coupling in a nonsymmorphic and noncentrosymmetric Kondo lattice model.
	We find that the combined time-reversal and inversion symmetry breaking associated with the Zeeman field and a sublattice-differentiating potential produces a progression of topological Lifshitz transitions.
	The Weyl nodes are annihilated at a Zeeman energy that is smaller than the Kondo energy scale and, thus, the strongly correlated topological semimetal retains its heavy fermion character under the magnetic field before it is quenched into a Kondo insulator.
	The complete node annihilation happens at a modest (laboratory-accessible) field. 
	 This reflects the extreme tunability of the strongly correlated topology to the Zeeman coupling, which is isolated from the tuning of the correlation physics {\it per se}.
	 The extreme topological tunability is also to be  contrasted with the case of weakly correlated Weyl semimetals where the Weyl nodes are hard to tune~\cite{Ong2016,Zhang2017,Caglieris2018,Ramshaw2018}. 	
	Our results are important for experiments on heavy fermion and related systems~\cite{dzsaber2019quenching}.
	In addition, our findings set the stage to map out a theoretical global phase diagram for strongly correlated topology. The latter
is important to the identification of new correlated topological states in theoretical models, and may guide systematic experimental explorations in this frontier area.

%%%%%%%%%%%%%%%%%%%%%%%%%%%%%%%%
\begin{figure*}[ht]
   \centering
	\includegraphics[width=2.07\columnwidth]{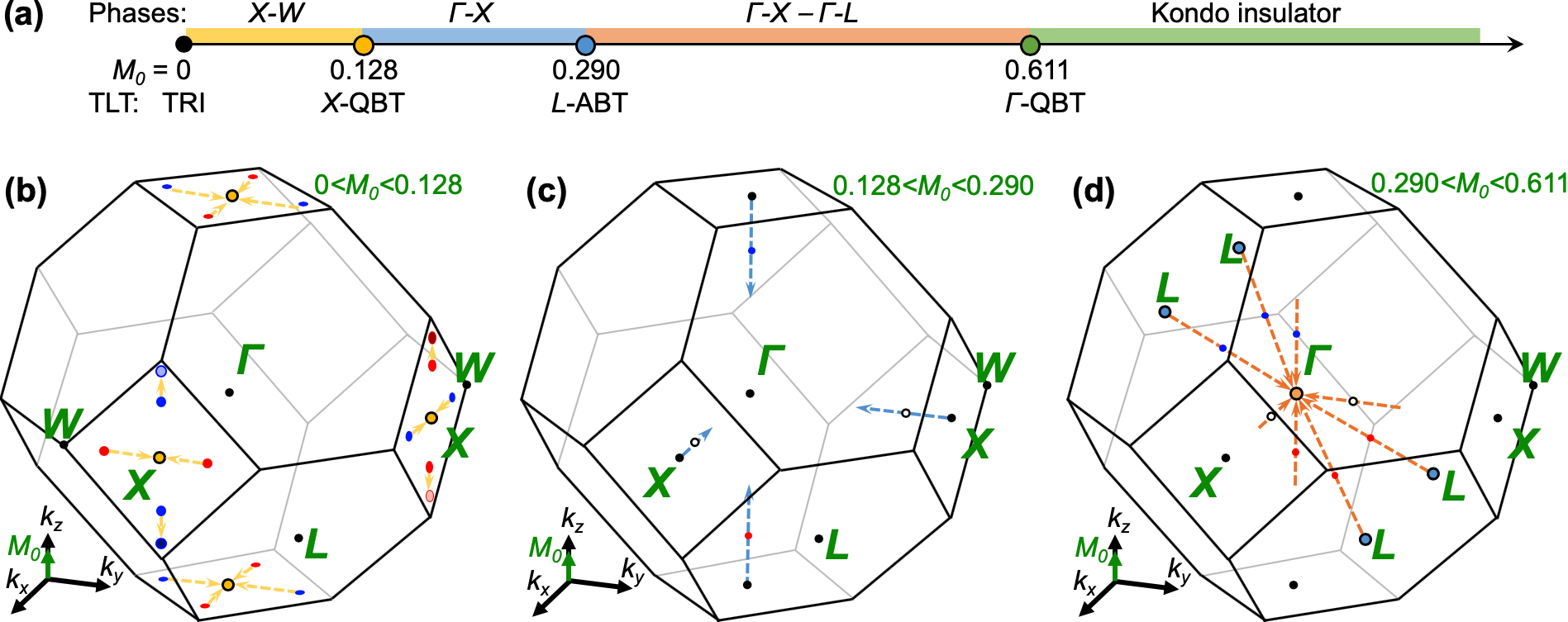}
   \caption{ \label{fig:phases-bz}
   (a) Phase diagram of the WKSM model as a function of the Zeeman field normalized by the zero-field Kondo temperature, $M_0=M_z/T_K^0$. 
   ``Phases'' labels each WKSM regime by its nodal trajectory in the BZ.
   TLT=topological Lifshitz transition, which labels regime crossover and critical points (filled circles) by where it occurs in the BZ and the type of dispersion.
   TRI=time-reversal invariant, QBT=quadratic band touching, ABT=anisotropic band touching.
   (b)-(d) shows the path of selected nodes through the fcc BZ for each corresponding WKSM regime: (b) $X$-$W$ (yellow), (c) $\Gamma$-$X$ (light blue), and (d) $\Gamma$-$X-\Gamma$-$L$ (orange).
	Red\,(blue) refers to $-1\,(+1)$ monopole Weyl nodes, darker\,(lighter) color variants denote nodes below\,(above) the Fermi energy, white-filled circles mark double Weyl nodes regime.
        }
\end{figure*}
%%%%%%%%%%%%%%%%%%%%%%%%%%	

%%%%%%%%%%%%%%%%%%%%%%%%%%%%%%%%%%%%%%%%%%%%%%%%%%%%%%%%%%%%%%%%%%%%%%%%%%%%%%%%%%%%%%%%%%

\textit{Model and Methods:}~~
	To consider the strong coupling (Kondo) limit, 
we implement the Anderson lattice model on a diamond crystal structure with
$\mathcal{H} = \mathcal{H}_c + \mathcal{H}_{cd} + \mathcal{H}_{d}$~\cite{WKSM_PRB2020,WKSM_PNAS}.
	The conduction electrons corresponding to the $spd$-orbitals of heavy fermion systems are represented by a Hamiltonian based on the Fu-Kane-Mele model~\cite{FKMmodel07,Ojanen13,Murakami2008},
\begin{align}
	\mathcal{H}_c 
	&= t\sum_{\la i j \ra, \sigma } \left( c_{i \sigma}^\dagger c_{j \sigma}+ \Hc \right) 
	+\sum_{i, \sigma}\left(m(-1)^i - \mu\right) n_{i \sigma}^c \nonumber\\
	&+ i \lambda \sum_{\la\la i j \ra\ra} \left[ c^\dagger_{i \sigma} \left( \bsigma\cdot {\bf e}_{ij} \right) c_{j\sigma} - \Hc\right],
	\label{ch5:eq:condel}
\end{align}
with nearest-neighbor $\la ij\ra$ hopping $t$, chemical potential $\mu$, a Dresselhaus-type spin orbit coupling $\lambda$, and an inversion symmetry breaking sublattice potential $m$.
	The electrons are coupled to the $d$ fermion species (representing physical $4f$ moments) through the hybridization term,
\begin{align}
	\mathcal{H}_{cd} &= V \sum_{i,\sigma} \left( d_{i \sigma}^\dagger c_{i \sigma} + \Hc\right).\label{ch5:eq:hybridization}
\end{align}
	The localized $4f$ electrons are represented by
\begin{align}
	\mathcal{H}_d&=\sum_i[ U n^d_{i \up} n^d_{i \dn} +
	\sum_{\sigma}(E_{d}\sigma_0+M_z\sigma_z)	
	d_{i \sigma}^\dagger d_{i \sigma}],\label{ch5:eq:hd}
\end{align}
where $E_d$ is much lower in energy than the conduction electron band edge~\cite{hewson_book}, $\sigma_0,\sigma_z$ are Pauli matrices acting on spin space, and the Coulomb repulsion $U$ acts between $d$ fermions. 
	The Zeeman interaction with field $M_z$ aligned in the $\hat{z}$ direction, breaks time-reversal symmetry.
	In a heavy fermion system subjected to a magnetic field, the response of the local moments dominates over orbital effects, so we do not consider Landau levels here. 
	
	In the $U \rightarrow \infty$ (Kondo) limit, we use the auxiliary boson method~\cite{hewson_book}, with $d_{i\sigma}=f_{i\sigma}b_i^\dagger$.
	Here, double-occupancy of the $d$-electrons is projected out, and small hole-fluctuations away from single-occupation are tracked by the bosonic field, $b_i\rightarrow \la b_i\ra=r$. 
	The occupation constraint is enforced by a term that replaces the Coulomb repulsion, 
	$\mathcal{H}_\ell=\ell\sum_{i,\sigma}(f^\dagger_{i\sigma}f_{i\sigma}+r^2-1)$.
	The  parameters $\ell$ and $r$, along with $\mu$, are determined by solving the system of saddle point equations $\la \delta\mathcal{H}/\delta r\ra=0,~\la \delta\mathcal{H}/\delta \ell\ra=0$, together with a condition for the electron filling.
	We consider a total quarter filling $n_d + n_c = 1$, which is $1$ fermion per site; in the strong coupling limit we have $n_f+r^2=1$, implying that $n_c\sim r^2$.
	
	Generically, both the time-reversal and inversion symmetry-breaking will influence the Weyl nodes~\cite{grefe2020_Zweyl}. 
	We study the extent to which the Kondo-driven Weyl nodal excitations survive the Zeeman coupling and, if they do, whether and how they can be controlled by the magnetic field.
	Our numerical method to solve the self consistency equations is detailed in the Supplemental Material~\cite{zeemansupp}.

%%%%%%%%%%%%%%%%%%%%%%%%%%%%%%%%%%%%%%%%%%%%%%%%%%%%%%%%%%%%%%%%%%%%%%%%%%%%%%%%%%%%%%%%%%

\textit{Zeeman-Field Tuning of the Weyl Nodes:}~~
	In what follows, we describe several Weyl node configurations in the Brillouin zone (BZ) that we encounter as a function of the Zeeman field. 
	We label each WKSM regime by the Weyl node trajectories from one high symmetry point to another, and label topological Lifshitz transitions (TLTs) between regimes/phases by their dispersion type and the BZ location, as shown in the phase diagram of Fig.~\ref{fig:phases-bz}(a).
	The Zeeman field strength is reported as $M_0=M_z/T_K^0$, relative to the Kondo temperature $T_K^0$, which is estimated from the bandwidth of the heavy Weyl bands for $M_z=0$, in units of $\mu_0=k_B=1$; 
	the parameters held constant are $\{E_d,V,\lambda,m,t\}=\{-7,9.29,0.5,1,1\}$.
	We previously established the WKSM phase for the time-reversal invariant case~\cite{WKSM_PRB2020,WKSM_PNAS}, represented by the black $M_0=0$ starting point of Fig.~\ref{fig:phases-bz}(a). 
	Here Weyl nodes can be found along all $X$-$W$ lines on the BZ boundary (red and blue dots of Fig.~\ref{fig:phases-bz}(b)).
	
	In the yellow $X$-$W$ regime, the Weyl nodes move along the yellow arrows in the BZ diagram in Fig.~\ref{fig:phases-bz}(b).	
	The darker\,(lighter) blue and red circles of nodes on the $k_x,k_y=\pm2\pi$ BZ faces (but off the $k_x$-$k_y-$plane) indicate that, in addition to moving in $\k-$space toward $W$, they shift in energy below\,(above) the Fermi energy ($E_F=0$), establishing Fermi/hole pockets that evolve with the field, which disappear in the $\G$-$X-\G$-$L$ regime (details in Ref.~\cite{zeemansupp}).
	Remarkably, all the nodes besides these (moving toward $X$) are pinned to $E_F$.
	In particular, the nodes remaining at $E_F$ are the $k_x,k_y=\pm2\pi$ BZ face nodes lying in the $k_x$-$k_y-$plane, and all four nodes on the $k_y=\pm2\pi$ faces.

	The $k_z=\pm2\pi$ face nodes meet uniformly at $X$, simultaneously with the pinned nodes on the $k_x,k_y=\pm2\pi$ BZ faces, while the off-$E_F$ nodes meet at $W$.	
	At $M_0=0.128$ (yellow circle, Fig.~\ref{fig:phases-bz}(a)), the dispersion forms a non-Kramers' type quadratic band touching at $X$, dubbed the $X$-QBT point.
	
%%%%%%%%%%%%%%%%%%%%
\textit{The emergence of double-Weyl nodes:}~~	
	Next, the WKSM enters the $\G$-$X$ regime (Fig.~\ref{fig:phases-bz}(c)).
	For $M_0$ just beyond the $X$-QBT point, the dispersion undergoes a TLT, and Weyl nodes form just inside the BZ boundary from the six $X$ points, and move towards $\G$ along the blue arrows.
	On the $k_x,k_y$-plane, double Weyl nodes develop with charge $+2$~\cite{Bernevig2012,Yang2014,Jian2015,Xiao2016,Huang2016,Ezawa2017,Ezawa2017R,Ahn2017,Yan2017,Zhang2018}, yet along the $\hat{z}$-axis, the Weyl nodes form a monopole charge-opposite pair.

%%%%%%%%%%%%%%%%%%%%%%%%%%
\begin{figure}[t]
   \centering
		\includegraphics[width=\columnwidth]{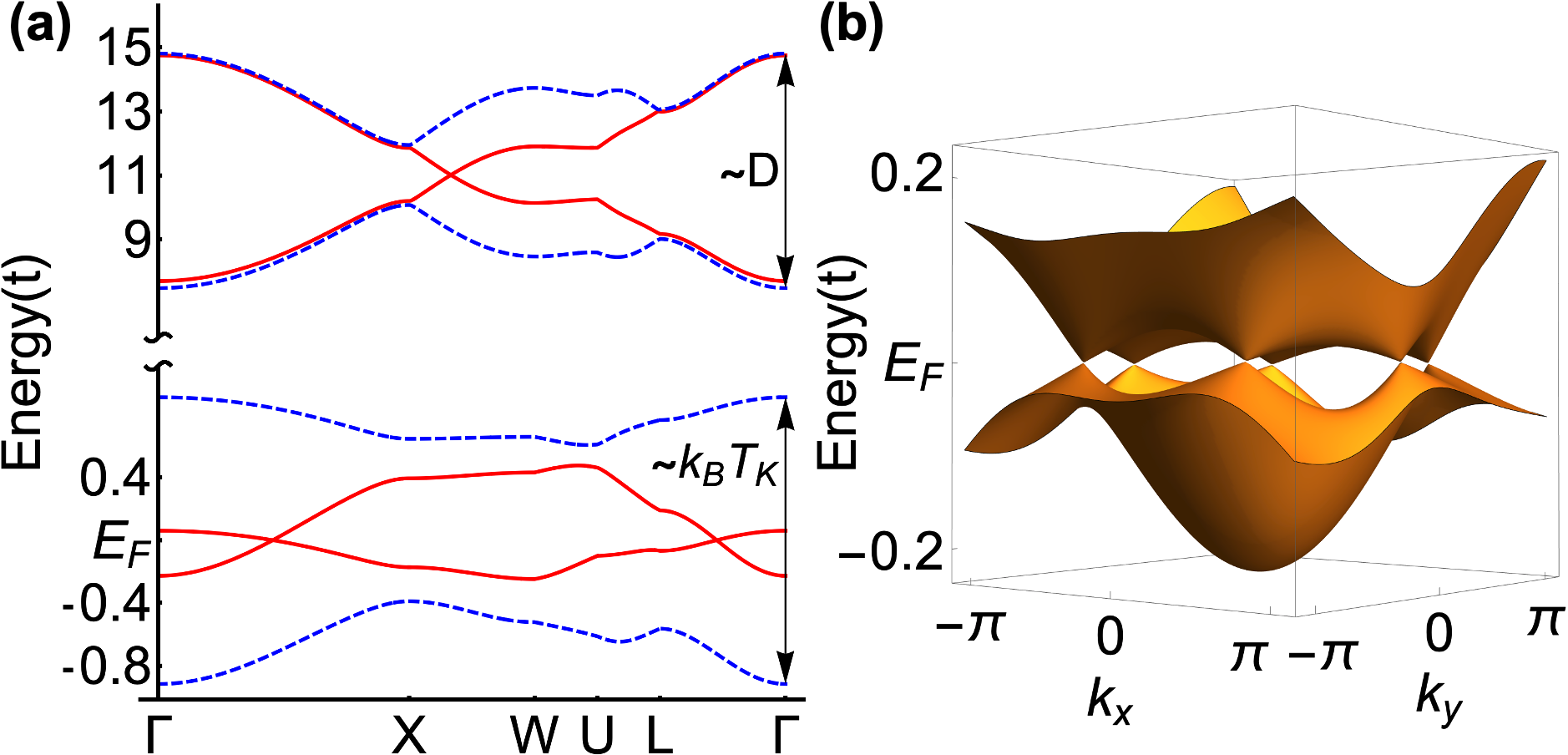}
   \caption{ \label{fig:mz-6}
   $\G$-$X-\G$-$L$ WKSM eigenenergy dispersion at $M_0=0.432$;
	(a) high symmetry path through the BZ,
    (b) on the $[011]$ plane, with origin at $\G$. 
    Here, $D$ is the bare conduction-electron bandwidth.
    }
\end{figure}
%%%%%%%%%%%%%%%%%%%%%%%%%%	

	At the $L$ points when $M_0=0.290$ (blue circles, Fig.~\ref{fig:phases-bz}(a),(d)), a crossover TLT occurs, called the $L$-ABT point to label an anisotropic band touching at $L$: the dispersion along the $(\pm1,\pm1,\pm1)$ directions is a quadratic band touching, but is linear in the perpendicular plane of the hexagonal BZ boundary.
	When $M_0>0.290$, charge $\pm1$ Weyl nodes form from the $L$ points along the $\G$-$L$ directions (see Fig.~\ref{fig:mz-6}).
	All sets of nodes (3 pairs of $\G$-$X$, 4 pairs of $\G$-$L$) progress towards $\G$, and this $\G$-$X-\G$-$L$ regime corresponds to the orange regime and arrows in Fig.~\ref{fig:phases-bz}(a),(d).

	%%%%%%%%%%%%%%%%%%%%%%%%%%
\begin{figure}[t]
   \centering
		\includegraphics[width=\columnwidth]{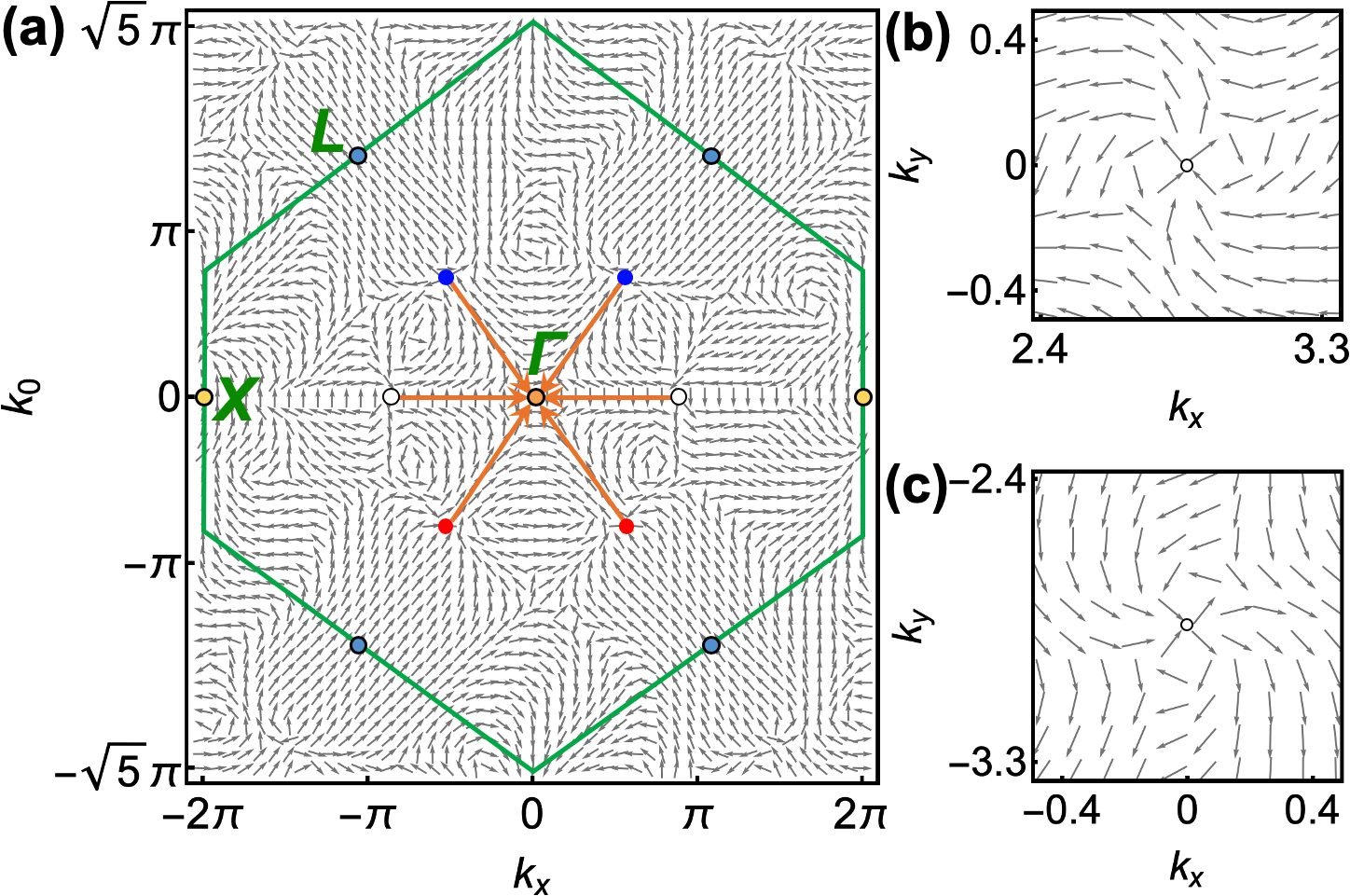}
   \caption{ \label{fig:ch5:berry}
   (a) Berry curvature projected onto the $[011]$ plane in the $\G$-$X-\G$-$L$ regime at $M_0=0.432$, where $k_0$ points along the $k_y$-$k_z-$plane. 
   The orange arrows indicate the node pairs' path toward simultaneous annihilation at the $\G$ point for $M_0>0.432$. 
   The green line marks the BZ boundary, blue/red\,(empty) circles mark $\pm1\,(+2)$ Weyl nodes, and yellow, blue, and orange circles correspond to the TLT locations of Fig.~\ref{fig:phases-bz}(a). 
   % need to add b and c description
    }
\end{figure}
%%%%%%%%%%%%%%%%%%%%%%%%%%
%	
	The normalized Berry curvature field of the $\G$-$X-\G$-$L$ WKSM is shown in Fig.~\ref{fig:ch5:berry}(a), projected onto the [011] plane to show both the $\G$-$X$ and $\G$-$L$ nodes.
	The $\G$-$L$ nodes (blue and red points) display sink/source monopole fields, while the $\G$-$X$ nodes have field structure for charge $+2$ double-Weyl fermions~\cite{Bernevig2012,Yang2014,Jian2015,Huang2016,Xiao2016,Ezawa2017,Ezawa2017R}.
	In Fig~\ref{fig:ch5:berry}(b)-(c), closer field configurations of the $\G$-$X$ double Weyl fermions are shown. 

	The emergence of the double Weyl fermions out of a quadratic band touching is reminiscent of what happens in diamond lattice $\alpha$-Sn~\cite{Barfuss2013,Rogalev2017}; in a $k\cdot p$ model, the application of a magnetic field to the pre-existing Luttinger semimetal leads to double-Weyl nodes at different energies~\cite{Zhang2018}. 
	There, it was found that a linear inversion symmetry breaking term splits the double-Weyl fermions into single-Weyl fermions. 
	By contrast, in our model the double-Weyl fermions are pinned to the Fermi energy, and are present over a range of combined time-reversal and inversion symmetry breaking couplings.
	The latter reflects the robustness of the Kondo effect that underlies the Weyl nodes.
	We also note that double Weyl fermions enhance the anomalous Hall effect twofold, and their Fermi arcs come in pairs~\cite{Ezawa2017}.

\textit{Zeeman-induced Annihilation of the Weyl Nodes:}~~
When the Zeeman coupling approaches the threshold value $M_0=0.611$, all the $\G$-$X$ and $\G$-$L$ nodes meet in the zone center $\G$. 
	This complete annihilation of the Weyl nodes by the Zeeman coupling is illustrated in Fig.\,\ref{fig:phases-bz}(d) as well as in Fig.\,\ref{fig:ch5:berry}(a).
 
\begin{figure}%[b!]
%[t]
   \centering
		\includegraphics[width=\columnwidth]{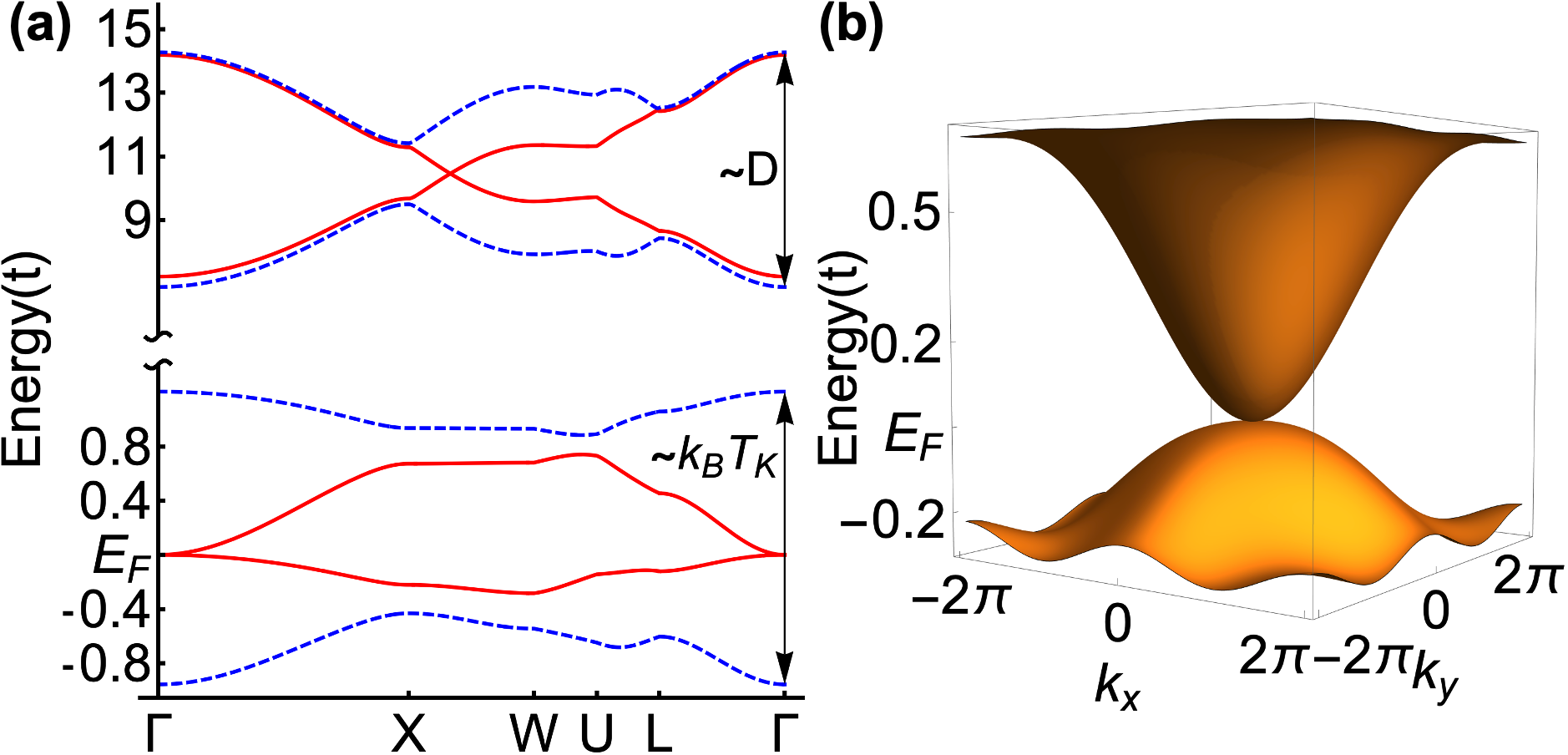}
   \caption{ \label{fig:mz-7}
	Eigenenergy dispersion at $\G$-QBT;
   (a) high symmetry path through the BZ,
   (b) on the $k_x$-$k_y$ plane, with origin at $\G$.}
\end{figure}

	At the threshold coupling $M_0=0.611$, a quadratic band touching critical point is formed (labeled $\G$-QBT) (orange circle, Fig.~\ref{fig:mz-7}).	
	The $\G$-QBT point is non-Kramers (singly degenerate) due to the broken time-reversal symmetry.
	When $M_0$ goes beyond this threshold, the $\G$-QBT bands open a gap, leading to a Kondo insulator phase.
	The emergence of the $\G$-QBT point is consistent with a continuous nature of the zero-temperature topological phase transition at this threshold $M_0$ value.
	
%
%%%%%%%%%%%%%%%%%%%%%%%%%%
%%%%%%%%%%%%%%%%%%%%%%%%%%

\begin{figure}[t]
   \centering
		\includegraphics[width=.55\columnwidth]{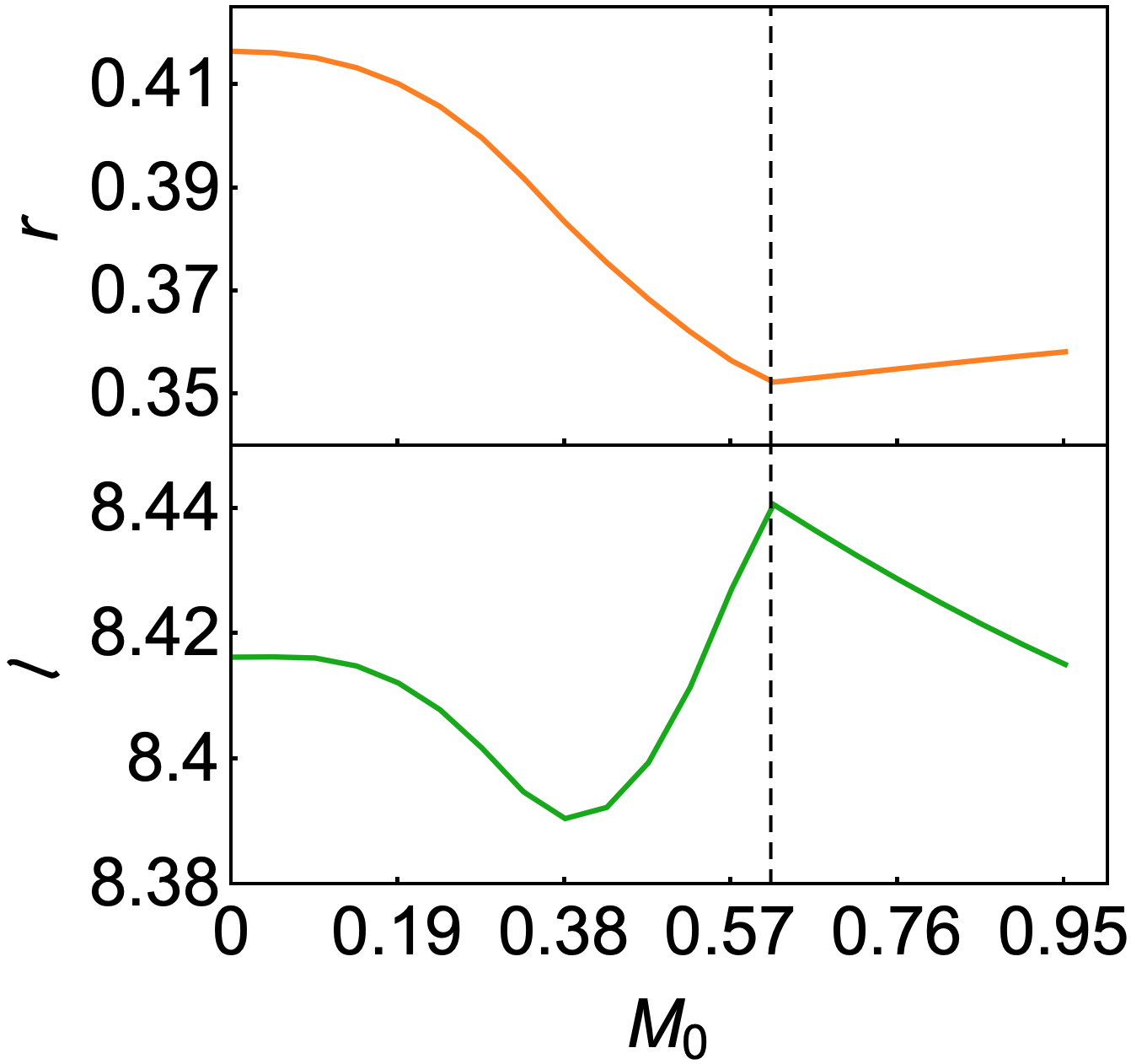}
   \caption{ \label{fig:ord-par}
   The saddle-point parameters as a function of $M_0$. 
   Top: scalar bosonic condensate $r$. 
   Bottom: Lagrange multiplier $\ell$.
   The dashed line marks the kinks at the $\G$-QBT point where $M_0=0.611$.
    }
\end{figure}

	Indeed, $\G$-QBT marks a second-order thermodynamic quantum phase transition.
	This is evidenced by the parameters $\ell$ and $r$ as a function of $M_0$ being continuous and showing nonanalyticities (pronounced kinks) across the $\G$-QBT point, as seen in Fig.~\ref{fig:ord-par}.

%%%%%%%%%%%%%%%%%%%%%%%%%%%%%%%%%%%%%%%%%%%%%%%%%%%%%%%%%%%%%%%%%%%%%%%%%%%%%%%%%%%%%%%%%%%%%%

\textit{Pinning of the Weyl nodes to the Fermi energy:}~~
	An important finding of our work is that each topologically nontrivial regime contains Fermi-energy-bound Weyl nodes.
	In the presence of time-reversal symmetry, this effect has been extensively demonstrated. 
	The strong correlation effect underlying the Kondo effect dictates that the Kondo-driven Weyl nodes lie near the Fermi energy within the narrow energy range of the Kondo scale; this effect, when combined with the nonsymmorphic nature of the space group symmetry, pins the Kondo-driven nodes in our model to the Fermi energy~\cite{WKSM_PNAS,WKSM_PRB2020}.
	The Zeeman field only reduces the symmetry of the spin degrees of freedom, and so the node-pinning mechanism remains free to operate.
	The space-group symmetry constraint is a general phenomenon~\cite{Cano2018,Watanabe2016,Kane_3ddirac}. 
	Thus, we expect that the mechanism we have advanced here, for the Zeeman-field-induced WKSM to Kondo-insulator transition, will apply to Kondo-lattice systems defined on a variety of nonsymmorphic crystalline structures.
		
	Indeed, we can see  the generality of the Fermi-level-bound nature of the Kondo-driven Weyl nodes by  comparing the Zeeman-tuned WKSM-Kondo insulator 
	transition to one that occurs as a function of tuning $m$ with time-reversal preserved~\cite{Ojanen13,WKSM_PNAS,WKSM_PRB2020} (even though the latter tuning is experimentally difficult to implement).
	When $m=0$, inversion symmetry is preserved, resulting in a Dirac-Kondo semimetal with degeneracies at $X$. 
	A WKSM forms for $0<m<2$ where the inversion-symmetry breaking splits the Dirac cone into two Weyl node pairs per square BZ boundary, with trajectories from $X$ to $W$.
	At $m=2$, each node meets its opposite charge partner from the neighboring BZ, becoming an anisotropic band touching at $W$, before annihilating and forming a gapped Kondo insulator state for $m>2$.
	Similar to the Zeeman-tuned WKSM-Kondo insulator transition, the merging of Weyl nodes at $W$ precipitates their annihilation and gap formation, which in both cases is accompanied by a kink in the saddle-point parameters across the critical point.
	Moreover, the topologically-nontrivial regime always has the Kondo-driven Weyl nodes at the Fermi energy.
	However, in the case of Zeeman tuning at nonzero $m$, the double Weyl nodes emerge, and the gap opening location occurs at the highly symmetric time-reversal-invariant momentum $\G$, in contrast to annihilation at the lower-symmetry time-reversal \emph{non}-invariant momentum $W$.

	%%%%%%%%%%%%%%%%%%%%%%%%%%
%%%%%%%%%%%%%%%%%%%%%%%%%%
%

%%%%%%%%%%%%%%%%%%%%%%%%%%	
\textit{Implications of our results:}~~
	Our results are highly relevant to the magnetic field experiments done on the nonmagnetic heavy fermion compound \cepd, where a WKSM-Kondo insulator transition has been observed~\cite{dzsaber2019quenching}.
	The experiments also found a heavy-fermion metal phase at high field values, which is more typical of Kondo-insulator physics and is not the focus of the present work.
	Importantly, we have shown that the WKSM phase and its TLTs take place at a Zeeman field smaller than the Kondo scale, before the Kondo effect itself is suppressed by the Zeeman coupling. 
	In other words, the extreme tuning of the strongly correlated topology happens when the strong correlation physics {\it per se} does not experience a qualitative change.

	A Zeeman coupling on the order of the Kondo scale in typical heavy fermion semimetals corresponds 
to a magnetic field on the order of 10 T. 
	In weakly-interacting WSM systems, such a magnetic field would have produced a minute Zeeman effect; indeed, typically, the orbital effect of the magnetic field dominates, which smears the Weyl fermions and prevent a well-controlled nodal annihilation~\cite{Zhang2017,Caglieris2018,Ramshaw2018,Ong2017}.
	We expect that the extreme topological tunability we have demonstrated also applies to related models based on Kondo effects~\cite{Feng13,Feng2016,Cook2017,Coleman_hwsm,Pixley2017,Cao2020}, as well as to other materials, such as \text{CeRu$_4$Sn$_6$}~\cite{Fuh2020,Paschen2010,Guritanu2013,Sunderman2015,Wissgott2016,Yu2016},
\text{YbBiPt}~\cite{Guo2018,Fisk1991,Chadov:2010aa,Mun2013} and
\text{CeSbTe}~\cite{Schoopeaar2317}, which are considered to be WKSM candidates.
	
\textit{Summary:}~~
This work has addressed the effect of Zeeman coupling in 
a nonsymmorphic and noncentrosymmetric Kondo lattice model, 
in which the cooperation of the Kondo effect and space-group symmetry produces 
Weyl nodes near the Fermi energy. 
	We have demonstrated an extreme responsiveness of the Weyl nodes to the Zeeman coupling.
	Several topologically-distinct semimetal regimes are induced by the Zeeman coupling, 
which involve double Weyl points that may significantly affect the anomalous magnetotransport properties.
	We have shown that a Zeeman coupling that is smaller 
than the zero-field Kondo energy scale is adequate to fully annihilate all the Weyl nodes, leading to a second-order topological quantum phase transition to a Kondo insulator. 
	Our results provide a proof-of-principle demonstration that the extreme tuning of strongly correlated topology can be realized without the interference of any qualitative change to the strong correlation physics {\it per se}.
	Equally important, our work sets the stage for the exploration of a global phase diagram for strongly correlated topology, which may be important for identifying new correlated topological states both theoretically and experimentally.

\begin{acknowledgements}
%%%%%%%%%%%%%%%%%%%%
We thank Pallab Goswami, Sami Dzsaber, Jennifer Cano, Diego Zocco, Mathieu Taupin for useful discussions.
Work at Rice has been supported by the NSF (DMR-1920740), the Robert A. Welch Foundation (C-1411) and the ARO (W911NF-14-1-0525). 
 Work at Los Alamos was carried out under the
auspices of the U.S. Department of Energy (DOE) National
Nuclear Security Administration under Contract No.
89233218CNA000001, and was supported by LANL LDRD Program.
Work in Vienna was supported by the Austrian Science Fund (projects P29279 and P29296) 
and the European Community (H2020 project 824109).

%%%%%%%%%%%%%%%%%%%%%%
\end{acknowledgements}
%%%%%%%%%%%%%%%%%%%%%%%%%%%%%%%%%%%%%%%%%%

%%%%%%%%%%%%%%%%%%%%%%%%%%%%%%%%%%%%%%%%%%%%%%%%%%%%%%%%%%%%%%%%%%%
\bibliography{/Users/sarahgrefe/master.bib}
%\bibliography{master.bib}
%%%%%%%%%%%%%%%%%%%%%%%%%%%%%%%%%%%%%%%%%%%%%%%%%%%%%%%%%%%%%%%%%%%
\end{document}